\documentclass[twocolumn,9pt]{article} 

\usepackage[square,numbers,sort&compress,comma]{natbib}

\usepackage{svg}
\usepackage{array}
\usepackage{amsmath}
\usepackage{amssymb}
\usepackage{caption}
\usepackage{graphicx}
\usepackage{latexsym}
\usepackage{stfloats}
\usepackage{times}
\usepackage[pagewise]{lineno}
\usepackage{hyperref}

\topmargin - 12pt 
\renewenvironment{abstract}%
              {
               \small
               {\bfseries \abstractname}
               \par
               \vspace{10pt}
              }

\renewcommand\abstractname{Abstract}

\newcommand{\nomenclature}
              [1]
              {
               \bgroup
               \flushleft
               \small\bf
               #1
               \par
               \egroup
              }

\renewcommand{\section}
              [1]
              {
               \bgroup
               \flushleft
               \small\bf
               \refstepcounter{section}
               \arabic{section}. #1
               \par
               \egroup
              }

\renewcommand{\subsection}
              [1]
              {
               \bgroup
               \flushleft
               \small\em
               \refstepcounter{subsection}
               \arabic{section}.
               \arabic{subsection}. #1
               \par
               \egroup
              }

\renewcommand{\subsubsection}
              [1]
              {
               \bgroup
               \flushleft
               \small\em
               \refstepcounter{subsubsection}
               \arabic{section}.
               \arabic{subsection}.
               \arabic{subsubsection}. #1
               \par
               \egroup
              }

  \newcommand{\acknowledgement}
              [1]
              {
               \bgroup
               \flushleft
               \small\bf
               #1
               \par
               \egroup
              }

  \newcommand{\sectionbib}
              [1]
              {
               \bgroup
               \flushleft
               \small\bf
               #1
               \par
               \egroup
              }

\begin{document}



\small
\baselineskip 10pt

\setcounter{page}{1}
\title{\LARGE \bf Effect of a magnetostatic field on laminar premixed hydrogen-air flames}

\author{
    {\large Tristan Lapaire$^{i,*}$, Sofiane Al Kassar$^{ii}$, Antonio Attili$^{ii}$}, Andrea Giusti$^{i}$\\[10pt]
    {\footnotesize \em $^i$Department of Mechanical Engineering, Imperial College London}\\[-5pt]
    {\footnotesize \em $^{ii}$Department of Multiscale Thermofluids, School of Engineering - The Edinburgh University}
}

\date{}  

\twocolumn[\begin{@twocolumnfalse}
\maketitle
\rule{\textwidth}{0.5pt}
\vspace{-5pt}

\begin{abstract} 

$\diamond$ \textit{This paper was first submitted in short version in November 2025, and in full version in March 2026}\\

Magnetic fields have shown potential to affect flame characteristics; however, the mechanisms of interaction are not fully understood. This paper investigates the effect of magnetic fields on premixed hydrogen-air flames that are prone to intrinsic instabilities, with a focus on the role of magnetic forces on the flame behaviour. The study is conducted using direct numerical simulations. Two flame conditions, both with an equivalence ratio of 0.5, are studied, one with the reactants at atmospheric conditions and the other at high pressure and high temperature. Different configurations of the magnetic field are investigated, each characterised by a different gradient of the square of the magnitude of the magnetic field, oriented in the direction opposite to the velocity of the incoming reactants. Results show that the investigated configurations of the magnetic field can reduce the flame consumption speed, an effect that is substantial in the lower pressure case, while it becomes negligible at high pressure. The effect of the magnetic forces increases with increasing gradient of the magnetic field and is mainly due to the reduction of the flame area. Results also show that the effects of magnetic fields on the reactivity of the flame and on the small cell structures developed along the flame front are negligible. Analysis of the force contributions demonstrates that the change in the flame area is caused by the rotational component of the magnetic forces, which alter the vorticity of the flow such that the finger-like structures formed by hydrodynamic instabilities tend to close. These forces are significant at low pressure, while they become negligible compared to the pressure gradient at high pressure. Ultimately, the results of this work indicate that magnetic forces have the potential to change the flame behaviour, a mechanism that could be used for active control of flame characteristics.

\end{abstract}

\vspace{10pt}

{\bf Novelty and significance statement} 

\vspace{10pt}
The effect of a magnetic field on the intrinsic instabilities of a laminar premixed hydrogen flame is investigated for the first time. The mechanism of interaction between the flame front and a magnetic field is revealed by decomposing the magnetic force into rotational and irrotational components. Rotational components of the forces and gradients in magnetic susceptibility of the mixture are identified as key elements for the effect of magnetic fields on the flame behaviour, which leads to an alteration of the flame area. The results of this work provide a new perspective into the action of magnetic forces on premixed flames, which opens up new possibilities for the development of technologies for the control of flame characteristics.

\vspace{5pt}
\parbox{1.0\textwidth}{\footnotesize {\em Keywords:} Hydrodynamic instability ; Thermodiffusive instability; Hydrogen; Magnetic force; Direct Numerical Simulation}
\rule{\textwidth}{0.5pt}
*Corresponding author.
\vspace{5pt}
\end{@twocolumnfalse}] 

\section{Introduction\label{sec:introduction}} \addvspace{10pt}

The transition to a low-carbon economy requires, among others, the identification of alternative energy carriers. Hydrogen, with its high specific energy and zero carbon emissions during combustion, is often presented as one of the most promising solutions to reduce emissions in sectors that are difficult to electrify. Despite its potential, direct combustion of hydrogen still faces many challenges, which are mainly related to the high reactivity of this fuel~\cite{academie_science_H2}.

In lean premixed hydrogen flames, combustion instabilities occur intrinsically. These instabilities appear to arise from two mechanisms. The first mechanism, usually referred to as hydrodynamic, or Darrieus–Landau, instability, originates from gas expansion across the flame front and results in a non-uniform flow velocity along the flame front, causing the growth of the flame-front perturbations~\cite{darrieus1938propagation, law2010combustion, matalon2007intrinsic, Matalon_2018}. The second mechanism is related to thermodiffusive instabilities, which arise from the strong differential diffusion caused by the non-unity Lewis number of hydrogen, promoting the formation of small-scale cellular structures and cusps on the flame surface~\cite{altantzis2012hydrodynamic, BERGER20191879, BERGER2022111935}. These complex flame front dynamics strongly modify the global flame properties~\cite{altantzis2012hydrodynamic, BERGER20191879, BERGER2022111935, BERGER2022111936}, with potential implications for the optimal and safe use of hydrogen combustion. Being able to control, or at least influence, these instabilities would therefore constitute a significant step towards the use of hydrogen in practical applications.

\begin{figure*}[ht!] 
  \centering
  \includegraphics[width=\textwidth]{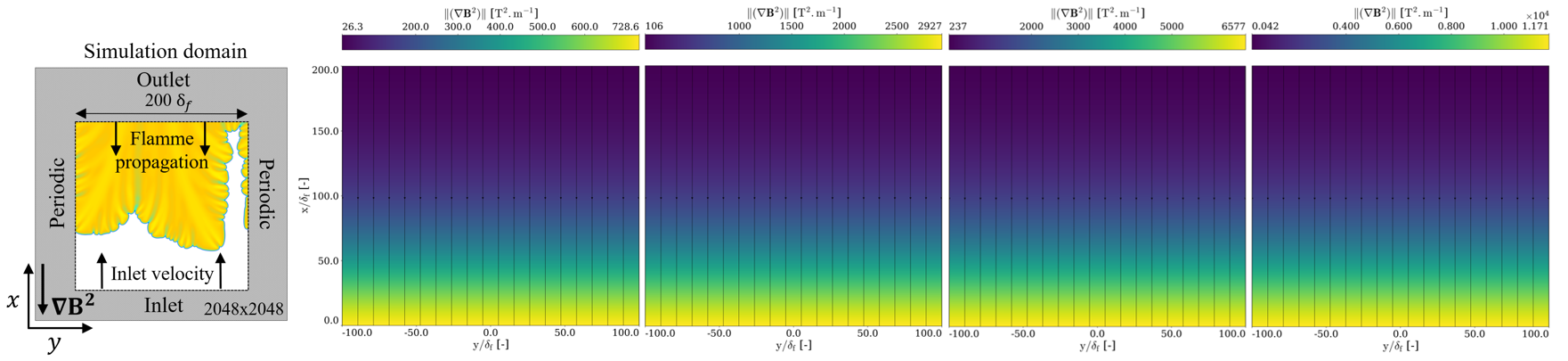}
  \caption{Simulation domain and gradient of the square of the magnetic field for configurations (b)–(e). The domain is 200 x 200 $\delta_f^2$, discretized on a 2048 x 2048 mesh ($\Delta \mathrm{x}$ = $\delta_f$ / 10.24 ). Both the flame propagation and the gradient of the square of the magnetic field are oriented in the negative $x$-direction.}
  \label{fig:mconfig}
\end{figure*}

In this context, magnetic fields could offer a means to control hydrogen flames through the magnetic forces acting on the species. The idea of such a control finds its origin in Faraday's pioneering work \cite{faraday1847lxiv}. Since then, the effect of magnetic fields on hydrogen flames has been investigated in laboratory flame configurations. Research findings suggest that the dominant response, arising a priori for non-uniform magnetic fields \cite{mizutani2001pre}, is primarily due to forces acting on paramagnetic species, in particular oxygen~\cite{wakayama1991behavior, wakayama1995magnetic, yamada2002numerical, yamada2003experimental}. These forces result in changes of the flame shape and distribution of species~\cite{yamada2003experimental,kinoshita2004numerical, hosseinpour2020computational,NATESH.2025}. However, these studies have mainly focused on the overall flame characteristics without providing a comprehensive analysis of the forces acting on the flame front and the effect of the characteristics of the magnetic field on the resulting flame dynamics. In addition, there are no detailed studies on the possible effect of an external magnetic field on the intrinsic instabilities of premixed hydrogen flames.

This study aims to investigate the interaction between magnetic forces and the flame front in laminar premixed hydrogen-air flames, with a focus on the effects of the magnetic field on the dynamics of the flame, including intrinsic instabilities, and overall flame characteristics. The specific objectives of the work are to: (i) investigate the impact of magnetic fields on the flame behaviour for increasing strength of the magnetic forces; (ii) study the effect of ambient conditions on the response of the flame to magnetic fields; (iii) identify the mechanism of interaction between magnetic forces and the flame front. To achieve these objectives, a premixed laminar flame with an equivalence ratio of 0.5 is investigated using direct numerical simulations under different pressure and temperature conditions and with different configurations of the magnetic field. 

\section{Methods \label{sec:confs&methods}} \addvspace{10pt}

\subsection{Configuration \label{subsec:conf}} \addvspace{10pt}

\begin{table}[b!]\footnotesize
\caption{Simulated cases and associated magnetic configurations (see Fig.~\ref{fig:mconfig}).}
\vspace{0.1cm}
\label{tab:cases_parameters}
\centering
\begin{tabular*}{\linewidth}{p{0.1\linewidth}ccc}
\hline
\noalign{\vspace{2pt}}
Case & \multicolumn{3}{l}{Thermodynamic parameters} \\
\hline
\noalign{\vspace{2pt}}
\centering 1. & \multicolumn{3}{l}{$p = 1$~atm ; $T_i = 298$~K ; $\mathrm{\phi = 0.5}$} \\
   & \multicolumn{3}{l}{$\delta_f \approx 4.57 \times 10^{-4}$~m} \\
   & \multicolumn{3}{l}{$\tau_f \approx 1.03 \times 10^{-3}$ s} \\
\noalign{\vspace{2pt}}
\centering 2. & \multicolumn{3}{l}{$p = 20$~atm ; $T_i = 700$~K ; $\mathrm{\phi = 0.5}$} \\
   & \multicolumn{3}{l}{$\delta_f \approx 1.89 \times 10^{-5}$~m} \\
   & \multicolumn{3}{l}{$\tau_f \approx 1.43 \times 10^{-5}$ s} \\
\noalign{\vspace{2pt}}
\hline
\noalign{\vspace{2pt}}
Config. & \multicolumn{3}{l}{$\| \mathrm{\nabla (B^2)} \|$, [T\textsuperscript{2}.m\textsuperscript{-1}]} \\
\noalign{\vspace{2pt}}
 & inlet & middle & outlet \tabularnewline
\hline
\noalign{\vspace{2pt}}
\centering (a) & $0$ & $0$ & $0$ \tabularnewline
\centering (b) & $7.28 \times 10^2$ & $1.52 \times 10^2$ & $2.60 \times 10^{1}$ \tabularnewline
\centering (c) & $2.93 \times 10^3$  & $6.13 \times 10^2$ & $1.10 \times 10^2$ \tabularnewline
\centering (d) & $6.58 \times 10^3$ & $1.37 \times 10^3$ & $2.37 \times 10^2$ \tabularnewline
\centering (e) & $1.17 \times 10^4$ & $2.45 \times 10^3$ & $4.20 \times 10^2$ \tabularnewline
\noalign{\vspace{2pt}}
\hline
\end{tabular*}
\end{table}

The configuration investigated in this work is a two-dimensional domain, schematically shown in Fig.~\ref{fig:mconfig}. The Cartesian frame of reference used throughout the study has origin at the centre of the inlet boundary. The premixed reactants enter the domain at $x=0$, with the flame propagating in the negative $x$-direction such that the flame is stabilised in the middle of the domain. Two different reactant conditions are investigated in this work: a low-pressure case with reactants at ambient conditions (Case 1), and a high-pressure condition with reactants at $T_i=700$~K and $p=20$~atm (Case 2). The equivalence ratio is set to $\phi = 0.5$ in both cases, which ensures the presence of both Darrieus-Landau and thermodiffusive instabilities~\cite{BERGER2022111936}. These conditions are summarised in Table~\ref{tab:cases_parameters}, together with the main characteristics of the magnetic field configuration investigated here. Five different configurations of the magnetic field are studied, each characterised by a different gradient of the square of the magnetic field, $\nabla(\mathrm{B^2})$. For reference, the four configurations where $\nabla(\mathrm{B^2})$ is non-zero are presented in Fig.~\ref{fig:mconfig}. In all configurations, $\nabla(\mathrm{B^2})$ is predominantly oriented in the negative $x$-direction, while the components of this gradient along $y$ are negligible, which make periodic boundaries applicable with very good approximation (note that, although negligible, these components are not zero and the resulting field is symmetric with respect to the $x$-axis). These magnetostatic fields have been obtained using Magpylib~\cite{ortner2020magpylib}, which allows us to obtain physically-consistent (i.e., divergence-free) magnetic fields. Note that the obtained configurations have a magnitude of $\nabla(\mathrm{B^2})$ that decreases with increasing distance from the inlet boundary (see Table~\ref{tab:cases_parameters}). The average magnitude of $\nabla(\mathrm{B^2})$ increases from configuration (a) to (e). Simulations of Case~1 and Case~2 were run for approximately 200 and 500 flame times, respectively, after having initialised them with perturbations~\cite{al2024efficient} to ensure the development of instabilities. 

\subsection{Governing equations and numerical methods \label{subsec:methods}} \addvspace{10pt}

The various cases are numerically studied using direct numerical simulation (DNS). The unsteady reacting Navier–Stokes equations are solved in the low Mach number limit, with the mixture obeying the ideal gas law \cite{lowmachNSl}. The transport equations for the species and temperature are solved using a finite-rate multistep chemical mechanism with 9 species and 46 reactions~\cite{Burke}. The viscosity of the mixture is calculated with Wilke’s formula \cite{wilke1950viscosity}. The transport properties are evaluated using a mixture-averaged model based on Bird’s expression \cite{bird2002transport, ATTILI2016192}. The diffusive fluxes of the species are evaluated using the Hirschfelder and Curtis approximation \cite{williams2018combustion, hirschfelder1964molecular}, with the thermodiffusive, or Soret, effect included using the Hirschfelder and Warnatz approximation  \cite{zirwes2025assessment}. The equations are solved with a semi-implicit finite difference code \cite{DESJARDINS20087125}, extensively used in previous work on reacting flows~\cite{BERGER20191879, NIEMIETZ20232209, al2024efficient}. In this work, the code has been further developed through the addition of magnetic force terms to the momentum and temperature transport equations, as well as to the diffusion of species. Spatial discretisation of the convective terms in the species transport equations is performed using a third-order WENO scheme~\cite{WENO}, while a second-order scheme is used for the diffusion terms and the momentum equation. The Strang operator splitting is applied to the chemical source term, which is integrated with the stiff ODE solver CVODE \cite{CVODE}.

In this study, only the magnetisation force is considered. This is expressed for each species $k$ using the Kelvin force density model \cite{MITelectromechanics}, which, developed at zero order using linear media assumption on the species, is given by
\begin{equation}
    \mathbf{f_k} = \frac{\chi_k}{2\mu_0} \nabla (\mathrm{B^2}), \label{eq:fK2}
\end{equation}
where $\mu_0$ is the magnetic permeability of free space and $\chi_k$ is the susceptibility of the $k$-th species, which represents the magnetic response of a species to a magnetic field. Note that the susceptibility of diamagnetic species is orders of magnitude lower than the susceptibility of paramagnetic species. Therefore, the magnetic force is only taken into account for paramagnetic species. For each species, the value of $\chi_k$ mainly depends on the local temperature and is expressed using the Curie law \cite{curie1895proprietes, physicalchemgordon} as

\begin{equation}\label{eq:chi}
    \chi_{k} = \frac{\rho Y_{k} \mathcal{N}_A g_{L_k}^2 \mu_B^2 \mu_0 \mathcal{S}_k(\mathcal{S}_k +1)}{3 W_kk_B T},
\end{equation}
where $\rho$ and $T$ are the local density and temperature of the mixture, respectively, $\mathcal{N}_A$ is the Avogadro number, $Y_k$ is the mass fraction of species $k$, $W_k$ is its molar mass, $\mu_B$ and $k_B$ are the Bohr magneton and Boltzman constant, respectively, $\mathcal{S}_k$ is total quantum spin and $g_{Lk}$ is the Landé g-factor~\cite{herzberg1944atomic}, approximated as 2. Values of $\mathcal{S}_k$ for the main paramagnetic species are summarised in Table~\ref{tab:SpinNumbers}. 

To provide insights into the mechanism of interaction between the magnetic field and the flame, the total force acting on the mixture, $\mathbf{f}_{\mathrm{KB^2}} = \Sigma_{k=1}^{N} \mathbf{f_{k}}$, is decomposed using the Helmholtz decomposition~\cite{batchelor2000introduction} as
\begin{equation}\label{eq:fhelmholtz}
    \mathbf{f}_{\mathrm{KB^2}} = \mathbf{f}_{\mathrm{KB^2}}^{\mathrm{irr}} + \mathbf{f}_{\mathrm{KB^2}}^{\mathrm{rot}}
\end{equation}

\noindent where $\nabla \times \mathbf{f}_{\mathrm{KB^2}}^{\mathrm{irr}} = \mathbf{0}$ and $\nabla \cdot \mathbf{f}_{\mathrm{KB^2}}^{\mathrm{rot}} = 0$. The irrotational part of the magnetic force, $\mathbf{f}_{\mathrm{KB^2}}^{\mathrm{irr}}$, can be expressed as the gradient of a scalar potential $\phi_{\mathrm{B}}$, such that $\mathbf{f}_{\mathrm{KB^2}}^{\mathrm{irr}} = \nabla \phi_\mathrm{B}$. For the zeroth-order approximation of the Kelvin force density, Eq.~\eqref{eq:fK2}, this decomposition allows the separation of the component of $\mathbf{f}_{\mathrm{KB^2}}$ that directly contributes to the pressure field, i.e.~the irrotational component of the force, from the component that drives rotational fluid motion.

\begin{table}[t!]\footnotesize
\caption{Total quantum spin numbers of the paramagnetic species in their fundamental state \cite{physicalchemgordon}.}
\vspace{0.1cm}
\label{tab:SpinNumbers}
\raggedleft
\begin{tabular}{p{0.12\linewidth}p{0.1\linewidth}p{0.1\linewidth}p{0.1\linewidth}p{0.1\linewidth}p{0.1\linewidth}}
\hline
Species & O$_2$ & O & H & OH & HO$_2$ \\
\hline
$\mathcal{S}_k$ & 1 & 1 & 1/2 & 1/2 & 1/2 \\
\hline
\end{tabular}
\end{table}

In order to analyse the spatial distribution of the various quantities of interest across the flame front, the progress variable $\mathrm{C_{H_2}}$ \cite{berger2023flame} is used, defined from the mass fraction of hydrogen, $Y_\mathrm{H_{2}}$, as  
\begin{equation}
    \mathrm{C_{H_2}} = 1 - Y_\mathrm{{H_{2}}} / Y_\mathrm{{H_{2,u}}},
\end{equation}
where $Y_\mathrm{{H_{2,u}}}$ is the hydrogen mass fraction in the unburnt mixture.

\newpage

\section{Results and discussion\label{sec:results}} \addvspace{10pt}

\begin{figure*}[t!] 
  \centering
  \includegraphics[width=\textwidth]{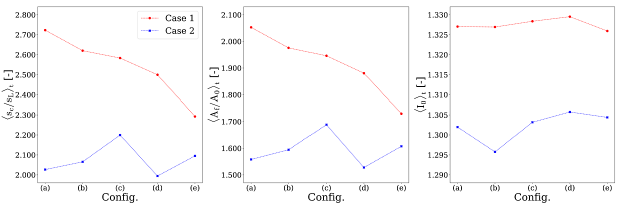}
  \caption{Time-averaged non-dimensional flame consumption speed (left), flame surface (middle), and stretch factor (right) for magnetic field configurations (a)–(e). The various quantities are computed as in \cite{berger2024effects} and averaged over 200 flame times for Case~1 and over the final 300 times for Case~2.}
  \label{fig:compaTall}
\end{figure*}

The global response of the flame to the different magnetic fields is investigated first. Figure~\ref{fig:compaTall} shows the three main flame dynamics quantities, i.e.~non-dimensional consumption speed, $s_c/s_\mathrm{L}$, non-dimensional flame surface, $A_f/A_0$, and stretch factor (usual notation applies), for all magnetic field configurations and both reactant conditions. In Case~1, the flame surface, $A_f/A_0$, decreases monotonically with increasing magnetic field strength, while $I_0$ remains approximately constant. This suggests that magnetic fields can affect the morphology of the flame by acting on the flame area. However, the local flame structure and chemistry are not significantly affected by the magnetic field, consistent with observations reported in~\cite{kinoshita2004numerical, hosseinpour2020computational}. Conversely, in Case~2, no clear trend is observed across all magnetic field configurations.

\begin{figure}[t!] 
  \centering
  \includegraphics[width=0.95\columnwidth]{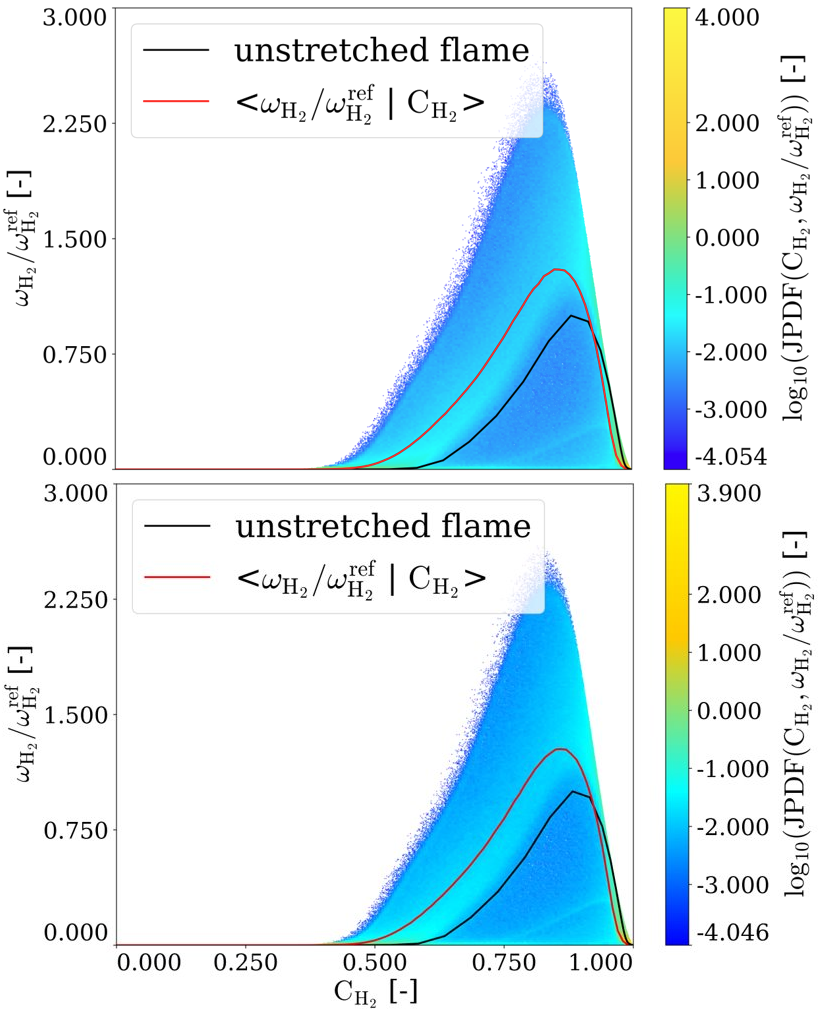}
  \caption{Logarithmic joint probability density function of the reduced reaction rate $\omega_{\mathrm{H_2}}/\omega_{\mathrm{H_2}}^{\mathrm{ref}}$ against the progress variable $\mathrm{C_{H_2}}$, for Case~1; magnetic field configuration~(a) on the top and (e) on the bottom. The red and black lines represent the conditional average of $\omega_{\mathrm{H_2}}/\omega_{\mathrm{H_2}}^{\mathrm{ref}}$ given $\mathrm{C_{H_2}}$, and the unstretched (one-dimensional) profile of $\omega_{\mathrm{H_2}}/\omega_{\mathrm{H_2}}^{\mathrm{ref}}$, respectively. The scaling is performed using the minimum reaction rate from the corresponding one-dimensional simulation.}
  \label{fig:reactionrate}
\end{figure}

This is further supported by Fig.~\ref{fig:reactionrate}, which shows the logarithmic joint probability density function of the reduced reaction rate as a function of the progress variable $\mathrm{C_{H_2}}$ for Case~1 and configurations~(a) and~(e) of the magnetic field. No significant changes in the distribution of the reaction rate across the flame front is observed between the two configurations, with the conditional averages remaining closely aligned across $\mathrm{C_{H_2}}$. Similar results are also found for Case~2 (not shown here). This confirms that the reaction rate distribution is unaffected by the magnetic field for all investigated cases.

\begin{figure}[t!] 
  \centering
  \includegraphics[width=0.82\columnwidth]{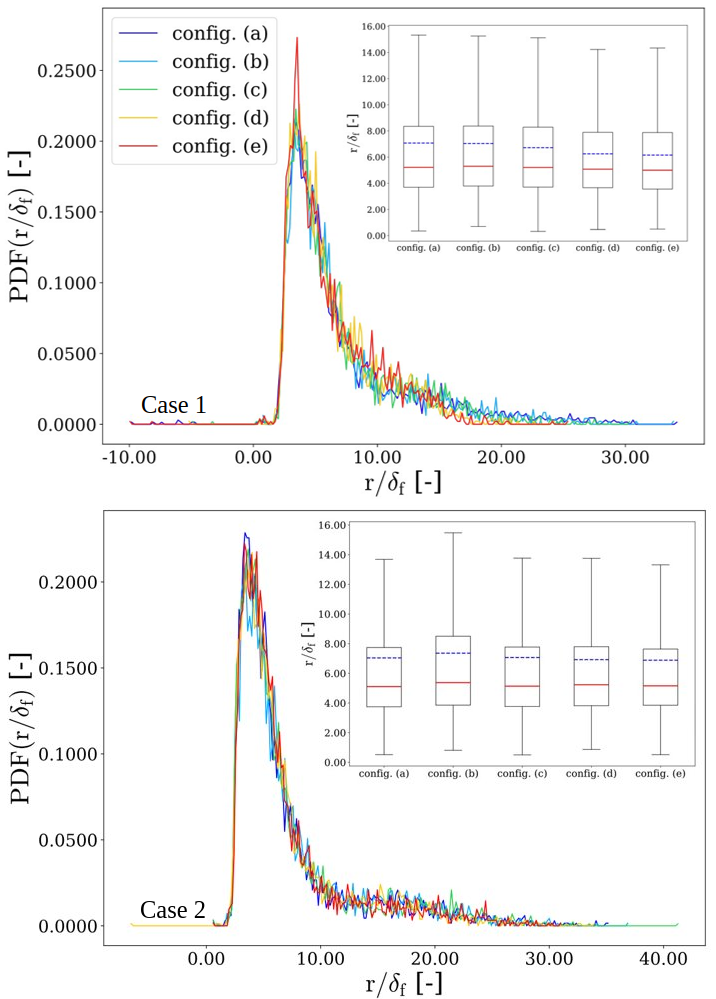}
  \caption{Probability density function of the reduced radius of the flame small-scale cellular structures for magnetic field configurations~(a)-(e), in Case~1 (top) and Case~2 (bottom). For each case, an inset box plot of the same quantity is shown, with the mean indicated by a dashed blue line and the median by a solid red line.}
  \label{fig:cellradius}
\end{figure}

An interesting metric for the evaluation of the local flame behaviour is the size of the flame small-scale cellular structures that develop on the flame front. Figure~\ref{fig:cellradius} shows the probability density functions of the reduced radius of the flame small-scale cellular structures for all investigated cases. The distributions are nearly identical across all configurations of the magnetic field for both reactant conditions. In Case~1, configuration~(e) of the magnetic field exhibits a slightly higher peak in the probability density function. However, the inset box plots in Fig.~\ref{fig:cellradius} show substantially equal mean and median values across all configurations, indicating that this deviation remains marginal and that the overall small-scale flame geometry is not significantly altered by the magnetic field. Together with Fig.~\ref{fig:reactionrate}, these results indicate that the magnetic field does not alter the local flame structure, its reactivity and its small-scale structures. Therefore, the thermodiffusive instabilities appear to remain unaffected. The reduction in $s_c/s_L$ observed in Fig.~\ref{fig:compaTall} for the case with reactants at ambient conditions is of purely fluid mechanics origin, as also suggested by the decrease in flame area. This indicates that the magnetic field affects primarily the hydrodynamic (Darrieus-Landau) instability.

\begin{figure}[t!] 
  \centering
  \includegraphics[width=0.9\columnwidth]{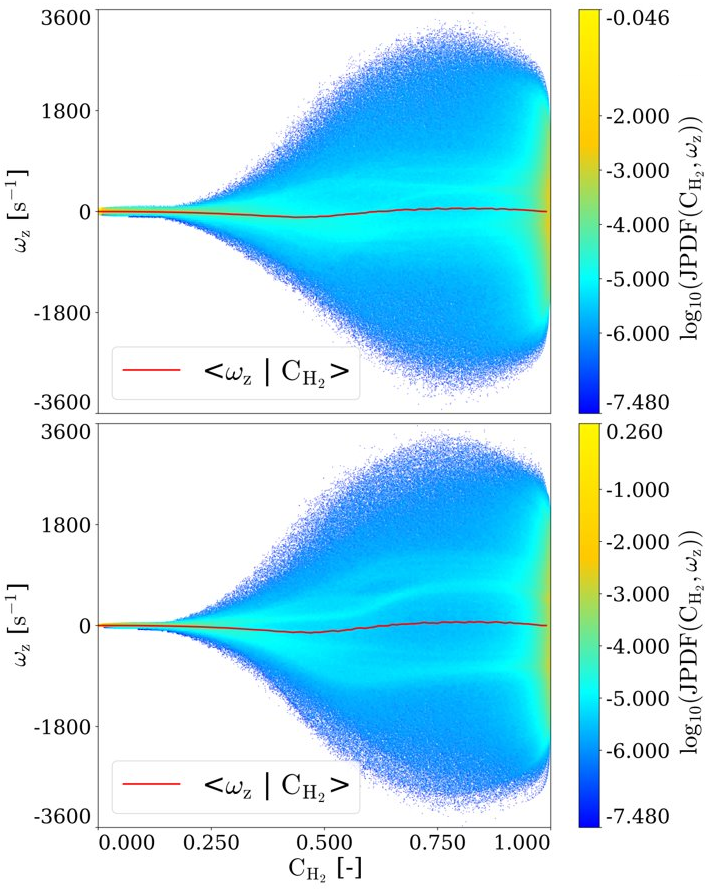}
  \caption{Logarithmic joint probability density functions of the $z$-component of the vorticity field $\omega_z$ against the progress variable, $\mathrm{C_{H_2}}$, for Case~1; magnetic field configuration~(a) on the top and (e) on the bottom. The red lines represent the conditional average of $\omega_z$ given $\mathrm{C_{H_2}}$.}
  \label{fig:omegaz}
\end{figure}

To investigate this hydrodynamic effect, Fig.~\ref{fig:omegaz} shows the vorticity distribution as a function of the progress variable $\mathrm{C_{H_2}}$ for Case~1 and the two extreme configurations of no magnetic field and magnetic field with the strongest gradient, i.e., configurations (a) and (e). Results show that the distribution of vorticity across the flame changes significantly when magnetic forces are introduced in the domain. The conditional average remains centred around zero, indicating no preferential rotation in the flow field on both sides of the flame front. However, in the presence of a magnetic field, the distribution of vorticity in the flame region ($\mathrm{C_{H_2}}>0.5$) exhibits two distinct peaks at positive and negative values of vorticity. Inspection of the results from all the cases indicates that the distance between these two peaks (i.e., stronger vorticity, both positive and negative) increases with increasing strength of the applied magnetic field gradient. This suggests that the magnetic field generates vortical structures that interact with the flame front. In the present configuration, these structures act to smooth the flame surface, resulting in a reduced flame surface area and consequently a reduced consumption speed, consistent with Fig.~\ref{fig:compaTall}. Note that in Case~2, no significant variation in the vorticity distribution is observed across all configurations (not shown here), consistent with the absence of clear effects on the flame behaviour.

\begin{figure}[t!] 
  \centering
  \includegraphics[width=0.75\columnwidth]{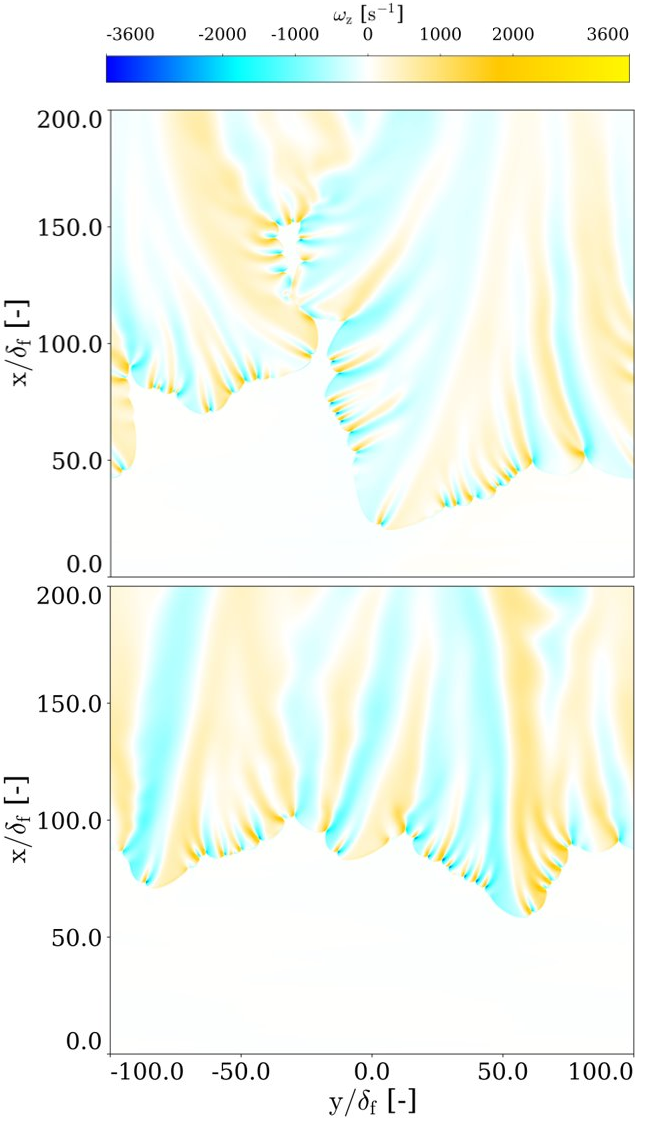}
  \caption{Instantaneous $z$-component of the vorticity field, $\omega_z$, at $t \approx 200\,\tau_f$, for Case~1; magnetic field configuration~(a) on the top and (e) on the bottom.}
  \label{fig:omegaz_case1}
\end{figure}

To further investigate the effect of magnetic forces on the flow field, the vorticity field in a selected time instant is shown in Fig.~\ref{fig:omegaz_case1}, for the two conditions analysed in Fig.~\ref{fig:omegaz}. At the same simulated time, the configuration with a magnetic field exhibits significantly higher vorticity magnitudes compared to the case with no magnetic field. In addition, it is evident that the size of the flame fingers, i.e. the Darrieus-Landau instability, is significantly reduced by the presence of a magnetic field, resulting in a smaller surface area. This structure of the vorticity field and its variation when a magnetic field is applied smooth the flame finger structures, driving the decrease in the flame surface observed in the time-averaged statistics.

\begin{figure*}[t!] 
  \centering
  \includegraphics[width=\textwidth]{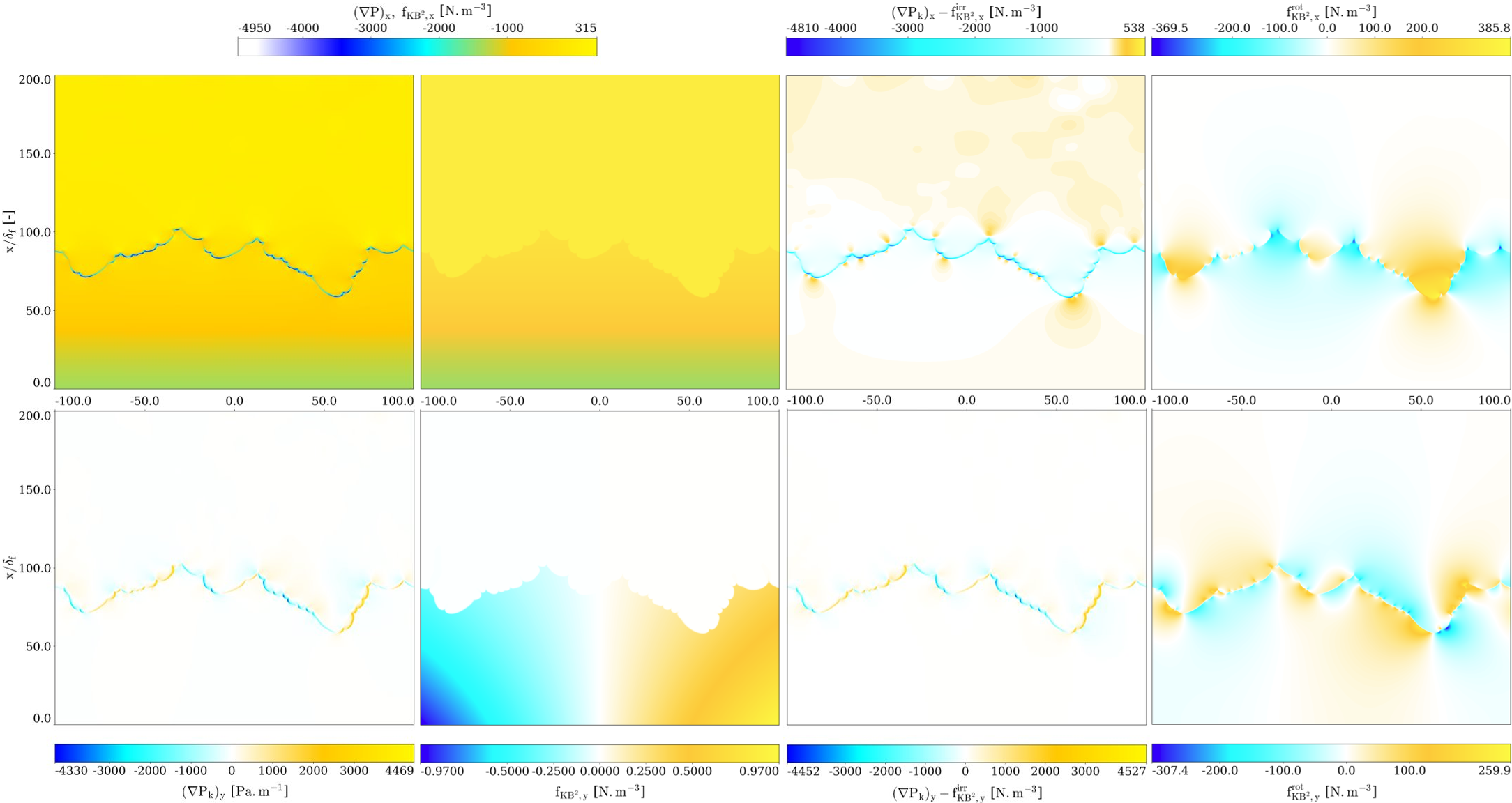}
  \caption{Instantaneous fields at $t \approx 200\,\tau_f$ for Case~1 and magnetic field configuration~(e); $x$-component (top row) and $y$-component (bottom row) of the pressure gradient $\nabla \mathrm{P_k}$, the total Kelvin force density $\mathbf{f}_{\mathrm{KB^2}}$, the difference $\nabla \mathrm{P_k} - \mathbf{f}_{\mathrm{KB^2}}^{\mathrm{irr}}$, and the rotational component $\mathbf{f}_{\mathrm{KB^2}}^{\mathrm{rot}}$.}
  \label{fig:case1forces}
\end{figure*}

To gain more insight into the mechanism leading to the hydrodynamic effects, the forces acting on the flow are discussed next. Figure~\ref{fig:case1forces} shows the $x$ and $y$-components of the pressure gradient and the Kelvin force for the ambient pressure condition (Case~1), for the magnetic field configuration with the strongest gradient. Magnetic force components resulting from the decomposition of the Kelvin force into irrotational and rotational parts are also shown. For the irrotational component of the magnetic force, the difference between the pressure gradient and this component is presented. Results demonstrate that the $x$-component of the pressure gradient, $(\nabla \mathrm{P_k})_x$, is significantly affected by the presence of the magnetic field. Close to the inlet, the values of the pressure gradient balance the Kelvin force, $\mathbf{f}_{\mathrm{KB^2},x}$. Both these forces exhibit negative values consistent with the orientation of $\nabla( \mathrm{B}^2)$ at the inlet. Note that $(\nabla \mathrm{P_k})_x$ displays slightly larger magnitudes than the magnetic forces in this region, consistently with evolving flow structures in the reactant region. As the flame front is approached, larger differences between the gradient of pressure and the Kelvin force are observed. The pressure gradient exhibits larger magnitudes along the flame surface, while the magnetic force transitions towards the values found on the product side of the flame without peaks on the flame front. Past the flame front, in the product region, both distributions recover a similar shape, with magnitudes of the magnetic force that tend to zero. This attenuation of the magnetic forces is primarily driven by the inverse dependence of the magnetic susceptibility on temperature. This is combined with the reduction of the mass fraction of oxygen in the product region, which is the species with the highest susceptibility and the one that contributes mostly to the overall magnetic force. 

Analysis of the irrotational and rotational components of the magnetic force shows that the difference $(\nabla \mathrm{P_k})_x - \mathbf{f}_{\mathrm{KB^2},x}^{\mathrm{irr}}$ is small in the reactant region and cancels on either side of the flame fingers in the vicinity of the flame front, with the exception of regions that show high flame curvature (see Fig.~\ref{fig:case1forces}, top). The rotational component of the magnetic force, $\mathbf{f}_{\mathrm{KB^2},x}^{\mathrm{rot}}$, pushes the flame fingers in the negative $x$-direction, thereby contributing to the flattening of the flame front and to the generation of vortical structures discussed previously. In the region immediately downstream of the flame fingers, the rotational term constitutes the sole effective force contribution, as $(\nabla \mathrm{P_k})_x$ and $\mathbf{f}_{\mathrm{KB^2},x}^{\mathrm{irr}}$ cancel each other, while on the flame front itself the pressure gradient remains the dominant term. Note that across the flame front, the flame is characterised by a gradient in density, which leads to an acceleration of the flow.

The contribution of forces is further analysed by considering the $y$-component of the forces (Fig.~\ref{fig:case1forces}, bottom). $(\nabla \mathrm{P_k})_y$ exhibits an antisymmetric pattern along the flame front, driven by the increase in temperature from reactants to products. The $y$-component of $\mathbf{f}_{\mathrm{KB^2}}$ is negligible, as the magnetic field gradient is predominantly oriented along $x$, although the symmetric values of the magnetic forces with respect to the $x$-axis remain visible in the force field. The difference $(\nabla \mathrm{P_k})_y - \mathbf{f}_{\mathrm{KB^2},y}^{\mathrm{irr}}$ is completely dominated by the pressure gradient, and $\mathbf{f}_{\mathrm{KB^2},y}^{\mathrm{rot}}$ 
displays a pattern consistent with a closing motion of the flame fingers, further supporting the action of the rotational forces towards reducing the flame area through the increase in vorticity. Note that in Case~2 (not shown here), $\nabla \mathrm{P_k}$ is orders of magnitude larger than the magnetic force for all magnetic field configurations, which supports the absence of significant effects of the magnetic field on the flame dynamics.

\begin{figure*}[t!] 
  \centering
  \includegraphics[width=\textwidth]{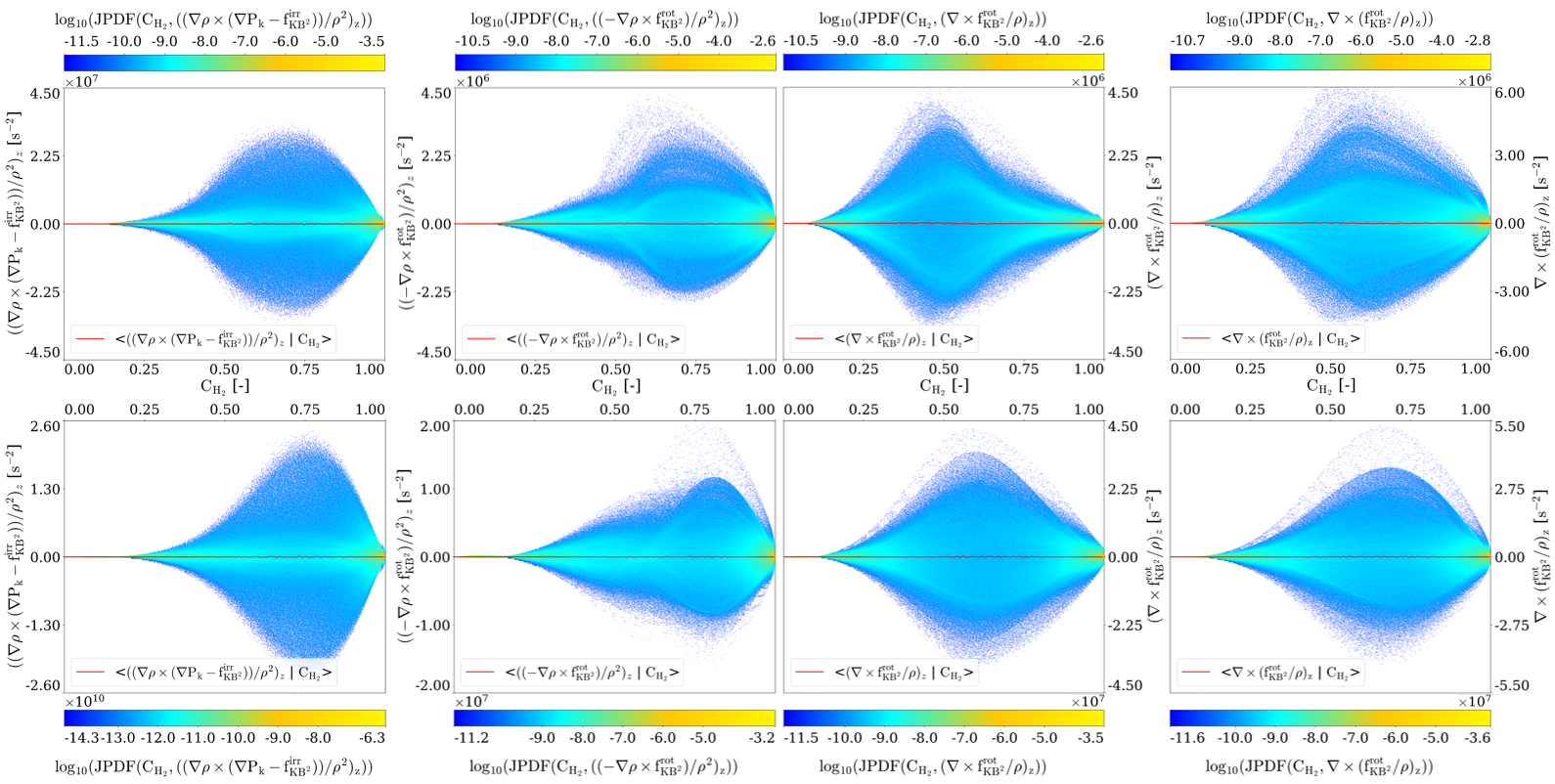}
  \caption{Logarithmic joint probability density functions of the various contributions to the $z$-component of the vorticity for Case~1 (top) and Case~2 (bottom) and the magnetic field configuration (e). From left to right, the contribution due to the difference between the pressure gradient and irrotational force, $\nabla \rho \times (\nabla \mathrm{P_k} - \mathbf{f}_{\mathrm{KB^2}}^{\mathrm{irr}}) / \rho^2$; the two contributions due to the rotational force, $- \nabla \rho \times \mathbf{f}_{\mathrm{KB^2}}^{\mathrm{rot}} / \rho^2$ and $\nabla \times \mathbf{f}_{\mathrm{KB^2}}^{\mathrm{rot}} / \rho$; the total contribution of the rotational force (sum of the two previous terms) $\nabla \times (\mathbf{f}_{\mathrm{KB^2}}^{\mathrm{rot}} / \rho)$, against the progress variable $\mathrm{C_{H_2}}$. The red lines represent the conditional average of the corresponding field given $\mathrm{C_{H_2}}$.}
  \label{fig:jpdfvorticiterms}
\end{figure*}

To investigate the mechanisms leading to the changes in vorticity, Fig.~\ref{fig:jpdfvorticiterms} shows the joint probability distributions of the source terms in the vorticity equation for Case~1 (top row) and Case~2 (bottom row) under the magnetic field configuration (e). The results demonstrate that, in both cases, the sources of vorticity peak in the flame region and that the term related to the pressure gradient is the dominant term in vorticity generation. When the term given by the difference between the pressure gradient and the irrotational component of the magnetic force is analysed (first column in Fig.~\ref{fig:jpdfvorticiterms}), results indicate that this contribution is only marginally affected by the magnetic field (other magnetic fields, not shown here, show a very similar distribution). The decomposition of the terms associated with the rotational part of the magnetic force (second and third columns in Fig.~\ref{fig:jpdfvorticiterms}) provides further insight into vorticity generation. The term in the second column is mainly driven by the density gradient, while the term in the third column represents the direct contribution of the rotational part of the magnetic force to the vorticity generation. In Case~1, these two terms are comparable and, when combined (last column in Fig.~\ref{fig:jpdfvorticiterms}), their magnitude becomes comparable to that of the pressure and irrotational term, leading to the changes in vorticity observed in Fig.~\ref{fig:omegaz}. Conversely, in Case~2, the values related to the pressure gradient are three orders of magnitude higher than the contributions given by the magnetic forces.

\begin{figure}[t!] 
  \centering
  \includegraphics[width=0.95\columnwidth]{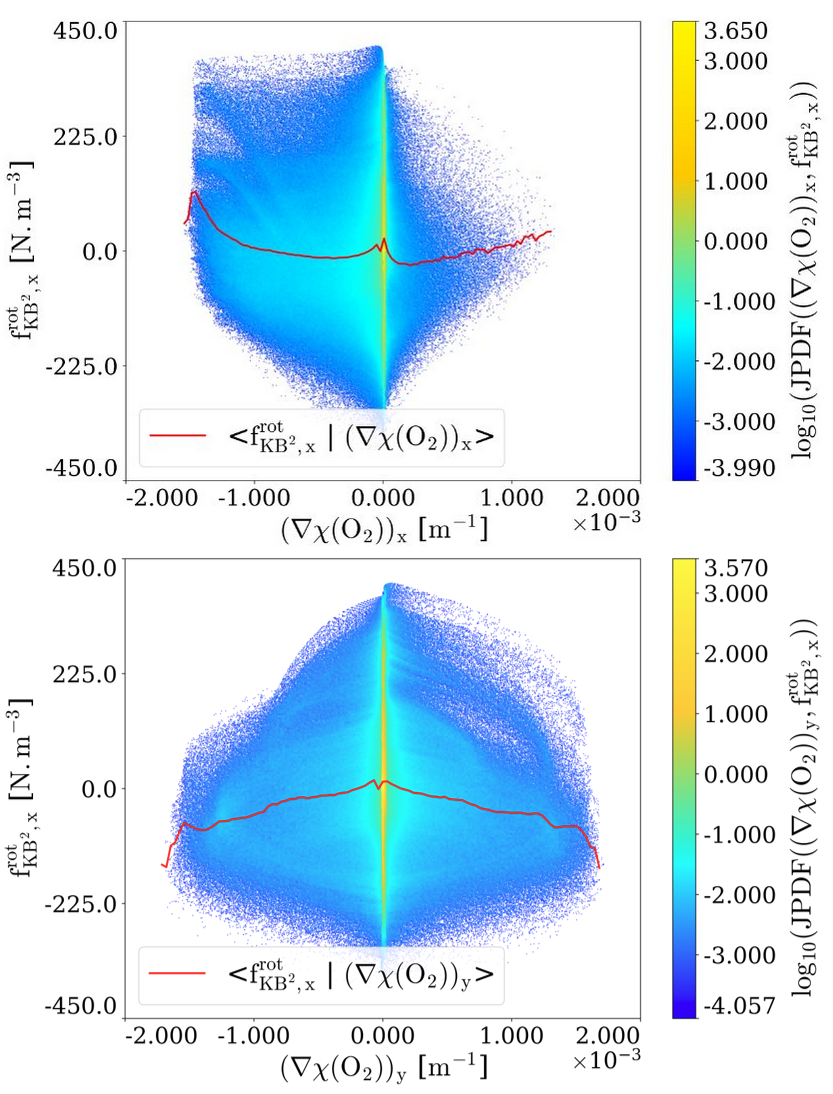}
  \caption{Logarithmic joint probability density function of the $x$-component of the rotational force against the $x$ (top) and $y$ (bottom) components of the oxygen susceptibility gradient, for Case~1, and magnetic field configuration~(e).}
  \label{fig:joint_susceptibility}
\end{figure}

It is important to note that the presence of rotational components of the magnetic force requires the presence of a gradient in susceptibility of the mixture. This could also be found through other decompositions of the magnetic force proposed in the literature. A common approach separates the Kelvin force density of Eq.~\eqref{eq:fK2} into a gradient of the so-called magnetically induced pressure and a magnetic Korteweg-Helmholtz term~\cite{Butcher_2023, MITelectromechanics}. In incompressible flows, the former component is irrotational and fully balanced by the pressure gradient~\cite{Butcher_2023}, similarly to the flame discussed in this study. Instead, the Korteweg-Helmholtz term, which is proportional to the gradient in susceptibility of the mixture, contains both irrotational and rotational contributions. Therefore, the presence of rotational components is expected to be linked to the presence of a gradient in susceptibility. Further inspection of the vorticity equation shows that vorticity is generated when a gradient in susceptibility is not aligned with the gradient of the magnitude of the magnetic field~\cite{Butcher_2023}. It is worth noting that rotational forces also appear in regions without significant gradients in mixture susceptibility, as a result of the force decomposition. The correlation between the gradient of susceptibility and the rotational component of the magnetic force for the cases investigated here is shown in Fig.~\ref{fig:joint_susceptibility} (the gradient in oxygen susceptibility is used). Results show that there is a peak in probability for zero values of the forces and zero gradient in susceptibility, which corresponds to the region far away from the flame. The joint probability density function between the $x$-component of the rotational force and the $y$-component of the susceptibility gradient is symmetric, indicating equal probability of having rotational forces on both sides of the fingers. Instead, the probability density function involving the gradient of susceptibility along $x$ is biased towards negative values of the susceptibility gradient, consistent with decreasing values of susceptibility from reactants to products. It is interesting to note that the maximum values of rotational forces are located in regions with negative values of the $x$-component of the gradient of susceptibility, while the $y$-component of the gradient of susceptibility for these peak values is around zero. Inspection of Fig.~\ref{fig:case1forces} shows that the peaks of rotational forces tend to appear at the apex of small structures formed on the flame front, which appear to be local maxima or minima of the flame surface, where the gradient of susceptibility changes sign in the $y$-direction. This phenomenon should be further investigated in future work. Extension of the present investigation to different equivalence ratios and three-dimensional domains, including more complex shapes of the magnetic field, as well as development of experiments to corroborate these findings, is also part of future research.

\section{Conclusions\label{sec:conlusion}} \addvspace{10pt}

The effect of magnetic forces on the dynamics of premixed laminar hydrogen-air flames has been investigated using direct numerical simulations for different conditions of the reactants and various strengths of the magnetic forces. Results show that at ambient pressure, magnetic forces affect the flame speed through changes in the flame area, whereas at high pressure, effects are negligible over the simulated flame time. A gradient in the magnetic field magnitude oriented opposite to the velocity of the incoming reactants leads to a decrease in the flame consumption speed, an effect that increases with increasing magnetic field gradients. The effect of magnetic fields on the reactivity of the flame is negligible at all conditions, as demonstrated by the statistics of the hydrogen reaction rate. Investigation of the small cell structures developed along the flame front also shows that magnetic fields do not alter their statistics, suggesting that the thermodiffusive instabilities remain unchanged. Investigation of the forces and their spatial distribution demonstrates that the observed changes in the flame area are due to the rotational component of the magnetic forces, which alter the vorticity of the flow such that the finger-like structures formed by Darrieus–Landau instabilities tend to reduce. These forces are significant at low pressure, while they become negligible compared to the pressure gradient at high pressure. The role of gradients in susceptibility in the establishment of rotational forces is also highlighted, providing further insight into the origin of the effect of magnetic forces on the flame behaviour. Ultimately, the results of this work indicate that magnetic forces have the potential to change the flame surface and stability, a mechanism that could be used to develop systems for active control of the flame behaviour and possibly inhibiting intrinsic instabilities of the flame.

\acknowledgement{CRediT authorship contribution statement} \addvspace{10pt}

\textbf{TL}: Conceptualization, Methodology, Validation, Investigation, Formal analysis, Visualization, Writing - Original Draft, Writing - Review \& Editing. \textbf{SAK}: Methodology, Writing - Review \& Editing. \textbf{AA}: Methodology, Writing - Review \& Editing. \textbf{AG}: Conceptualization, Formal analysis, Writing - Original Draft, Writing - Review \& Editing.

\acknowledgement{Declaration of competing interest} \addvspace{10pt}

The authors declare that they have no known competing financial interests or personal relationships that could have appeared to influence the work reported in this paper.

\acknowledgement{Acknowledgments} \addvspace{10pt}
TL and AG acknowledge financial support from UKRI EPSRC grant ref: EP/Y031423/1, which is part of the ICHAruS project, funded by the European Union under the Horizon Europe program, grant n.~101120321.

\footnotesize
\baselineskip 9pt

\thispagestyle{empty}
\bibliographystyle{proci}
\bibliography{references.bib}


\newpage

\small
\baselineskip 10pt


\end{document}